\documentstyle[ams]{article}

\def\beq{\begin{equation}}
\def\eeq{\end{equation}}
\def\bea{\begin{eqnarray}}
\def\eea{\end{eqnarray}}
\def\nn{\nonumber}
\begin{document}

\catcode`\@=11
\def\unbracefonts@{\relax}
\catcode`\@=12

\begin{titlepage}
\begin{center}

{{\it P.N.Lebedev Institute preprint} \hfill FIAN/TD/05-91\\
{\it I.E.Tamm Theory Department} \hfill ITEP-M-5/91
\begin{flushright}{August 1991}\end{flushright}
\vspace{0.1in}{\Large\bf  From Virasoro Constraints in Kontsevich's Model\\
 to $\cal W$-constraints in 2-matrix Models}\\[.4in]
\large  A.Marshakov, A.Mironov}\\
\bigskip {\it Department of Theoretical Physics \\  P.N.Lebedev Physical
Institute \\ Leninsky prospect, 53, Moscow, 117 924}
\footnote{E-mail address: theordep@sci.fian.msk.su},\\ \smallskip
\bigskip {\large A.Morozov}\\
 \bigskip {\it Institute of Theoretical and Experimental
Physics,  \\
 Bol.Cheremushkinskaya st., 25, Moscow, 117 259}
\end{center}
\bigskip
\bigskip

\centerline{\bf ABSTRACT}
\begin{quotation}

The Ward identities in Kontsevich-like 1-matrix models are used to prove at the
level of discrete matrix models the suggestion of Gava and Narain, which
relates the degree of potential in asymmetric 2-matrix model to the form of
$\cal W$-constraints imposed on its partition function.
\end{quotation}
\end{titlepage}

\setcounter{page}1

\section{Introduction}

   Matrix models, originally developed [1] as an alternative approach to (so
far
2-dimensional) quantum gravity, are now understood to possess a deep and
interesting mathematical structure of their own. Indeed, their partition
functions are usually subjected to infinite sets of constraints, which can
be formulated in a form of differential equations with respect to the
 "time"-variables, and as a corollary they
appear proportional to "$\tau$-functions" of integrable hierarchies [2].

   While this general scheme is more or less well understood at the level of
discrete 1-matrix models [3,4], it is still in many respects an open question
when continuum limit and/or {\it multi}matrix models are concerned. An
 important
breakthrough in the study of continuum limit of 1-matrix models is due to
a recent conjecture of M.Kontsevich [5], who in fact suggested that the
proper continuum limit of the Hermitean 1-matrix model can be described in
terms of a somewhat different matrix model, to be referred later as
Kontsevich's model. Its partition function is essentially proportional to
the  $N \times N$ (anti)Hermitean matrix integral

\beq
{\cal F}\{\Lambda \} \equiv  \int   DX\ \exp(- trW[X] + tr\Lambda X)
\eeq
with  $W[X] = X^3$  and  $N \longrightarrow \infty$. This conjecture was
 strongly supported by
a recent proof [6] that partition function of Kontsevich's model satisfies
exactly the same set of differential equations (Virasoro constraints [2]) as
continuum limit of the original Hermitean 1-matrix model (so that it
remains only to understand what are the requirements, which guarantee the
uniqueness of the solution of these Virasoro constraints). In [1]
it was shown that Virasoro constraints arise from the obvious Ward identities,
satisfied by ${\cal F}(\Lambda)$ ,

\beq
{(tr \Lambda ^pW'[\partial /\partial \Lambda _{tr}] -
tr\ \Lambda ^{p+1}){\cal F}\{\Lambda \} = 0}
\eeq
as a result of a change (Miwa transformation) of the argument

\beq
\Lambda   \rightarrow   \{T_m = {1\over m+{1\over 2}} tr\Lambda ^{-m-{1\over2}}
- {4\over 3\sqrt{3}}\delta _{m,1}\} \hbox{   .  }
\eeq
In its turn, eq.(2) states that the integral (1) is invariant under
infinitesimal shift of integration variable,

\beq
X  \rightarrow   X + \Lambda ^p \hbox{  .  }
\eeq
The implication of Kontsevich's conjecture in this context (for more
physical implication see ref.[6]) is that the Miwa transformation (3)
allows us to substitute a sophisticated double-scaling limit in
conventional matrix models [1] by a naive limit of  $N \longrightarrow \infty$
   in
the model like (1). This opens promising possibilities in the study
of continuum matrix models.

   Somewhat unexpectedly this fresh view on continuum limits provides
also a new approach to the study of {\it discrete} multimatrix models. One
of the main problems about them is the lack of understanding (and
even a derivation) of differential equations which substitute Virasoro
constraints of 1-matrix models (and are believed [2] to be expressible in
terms of generators of $\cal W$-algebras).
 What we are going to demonstrate in this letter
is that just the same Ward identities (2) for the matrix integral (1)
provide a simple proof of the $\cal W$-like constraints in a discrete (i.e.
at finite $N$) 2-matrix model.

Moreover, the recent observation due to E.Gava and K.Narain [7], stating that
the spin of $\cal W$-constraint in fact coincides with the power of potential
$W[X]$ is naturally explained in this way. (This suggestive result has been
obtained in [7] by a tedious examination of continuum limit of loop equations
for specific 2-matrix model with  $W[X] = X^3).$

\section{The main results}

The partition function of the discrete Hermitean 2-matrix model [8] is given by
a double integral over  $N\times N$  Hermitean matrices  $X$  and  $\Lambda :$

\beq
Z_{V,W} = \int   DX\ D\Lambda \ \exp ( - tr\{V[\Lambda ] + W[X] - \Lambda X\})
\equiv \int   D\Lambda \ e^{- trV[\Lambda ]}{\cal F}_W\{\Lambda \} \hbox{   .
}
\eeq
The potentials  $V$  and $W$  are conventionally parameterized by the
corresponding
time-variables

\bea
V[\Lambda ] = \sum _{k\geq 0}^{\infty}t_k\Lambda ^k \hbox{  ,  }
\nn \\
W[X] = \sum _{k\geq 0}^K s_kX^k
\eea
and the partition function  $Z_{V,W}$ is usually treated as a functional of
$\{t_k\}$  and  $\{s_k\}$. Below we will use the obvious notation  $Z_{V,W}
\equiv  Z_W\{t_k\}$. While in [2] it was suggested that the continuum $(i.e$.
$N \rightarrow  \infty )$ limit of  $Z_{V,W}$ in the case of  $W = V$ (this $K=
\infty$) is
annihilated by a set of operators which form  ${\cal W}_3$-algebra, the results
of ref.[7] imply, at least, that the structure of constraints in the asymmetric
situation  $W \neq  V$  is more complicated: generators of
${\cal W}_K$-algebra (expressed in terms of  $t$-variables) annihilate
$Z_W\{t_k\}$  whenever  $W[X]$  is a polynomial of power  $K.$
Our purpose below is to explore the origin of this important phenomenon in the
most transparent way.

The natural derivation arises from comparison of eqs. (5) and
(2). Indeed:

\beq
{\partial Z_W\over \partial t_{p+1}} {=}
 - \int   D\Lambda \ e^{- trV[\Lambda ]} tr\Lambda ^{p+1} {\cal F}\{\Lambda \}
 {=}
- \int   D\Lambda \ e^{- trV[\Lambda ]}
tr(\Lambda ^p W'[{\partial \over \partial \Lambda _{tr}}])
 {\cal F}\{\Lambda \} \hbox{  .  }
\eeq
After integration by parts it turns into

\beq
\int D\Lambda {\cal F}\{\Lambda \} tr(W'[-{\partial
 \over \partial \Lambda _{tr}}] \Lambda ^p) e^{- trV[\Lambda ]}
= \sum ^K_{k>0}ks_k \int D\Lambda {\cal F}\{\Lambda \}
tr((-{\partial \over \partial \Lambda _{tr}} )^{k-1} \Lambda ^p)e^{-
trV[\Lambda ]} \hbox{  .  }
\eeq
The leading term in the sum on the r.h.s., i.e.

\bea
Ks_K \int D\Lambda {\cal F}\{\Lambda \}
\left\lbrace tr\ \Lambda ^p(V'[\Lambda ])^{K-1} +
O(V^{K-2})\right\rbrace = \nn \\
= Ks_K \sum _{a_1,...,a_{K-1}}a_1t_{a_1}\ldots a_{K-1}t_{a_{K-1}}
{\partial \over \partial t_{a_1+...+a_{K-1}+p+1-K}} Z_W\{t\}
\eea
is a ``classical" \footnote{In the sense of ref.[9]}
piece of the operator  ${\cal W}^{(K)}_{p+1-K}$ --- ($p+K-1$)-th
harmonic of the spin-$K$ generator of  ${\cal W}_K$-algebra --- acting on
$Z_W\{t\}$. (Note also that according to (8)  $p\geq 0)$. This explains
essentially the very origin of  ${\cal W}_K$-constraint and its intimate
relation to the form of potential  $W[\Lambda \}$  in complete agreement with
[7].

In general, the Ward identity (7) can be rewritten as a set
of constraints

\beq
\left( - {\partial \over \partial t_{p+1}} +
\sum _{k>0}^K ks_k\tilde {\cal W}^{(k)}_{p+1-k}\{t\}\right) Z_W\{t\} = 0
\hbox{  .  }
\eeq
Operators $\tilde {\cal W}^{(k)}$ are defined by

\beq
  \tilde {\cal W}^{(k)}_{p+1-k}
e^{-trV[\Lambda ]} = - tr((-{\partial \over \partial \Lambda _{tr}} )^{k-1}
 \Lambda ^p)e^{-
trV[\Lambda ]} \hbox{  .  }
\eeq
 They obey recurrent relation

\beq
\tilde {\cal W}^{(k+1)}_p = \sum _n nt_n \tilde {\cal W}^{(k)}_{n+p} +
\sum _{a+b=p+k-1} {\partial \over \partial t_a} \tilde {\cal W}^{(k)}_{b+1-k}
\hbox{   ;  } p \ge -k
\eeq
with

\beq
\tilde {\cal W}^{(2)}_p = {\cal L}_p = \sum _n nt_n {\partial \over \partial
t_{n+p}}
+ \sum _{a+b=p} {\partial ^2 \over \partial t_a \partial t_b}
\hbox{   ;  } p \ge -1 \hbox{  ,  }
\eeq

\beq
\tilde {\cal W}^{(1)}_p = {\cal J}_p = {\partial \over \partial t_p}
\hbox{  ;  } p \ge 0 \hbox{  .  }
\eeq
In order to derive (12) one has to take explicitly one
derivative $\partial / {\partial \Lambda_{tr}}$ in (11):

\bea
\lefteqn{\tilde {\cal W}^{(k+1)}_{p}e^{-trV[\Lambda ]} = } \nn \\
& & = tr((-{\partial
\over \partial \Lambda _{tr}} )^{k}\Lambda ^{p+k})e^{-trV[\Lambda ]} =
tr((-{\partial \over \partial \Lambda _{tr}} )^{k-1}(-{\partial \over
\partial \Lambda _{tr}}) \Lambda ^{p+k})e^{-trV[\Lambda ]} = \nn \\
& & = tr((-{\partial \over \partial \Lambda _{tr}} )^{k-1} V'(\Lambda )
\Lambda ^{p+k})e^{-trV[\Lambda ]} - \sum _{a+b=p+k-1}
tr((-{\partial \over \partial \Lambda _{tr}} )^{k-1}
\Lambda ^b) tr \Lambda ^a e^{-trV[\Lambda ]} = \nn \\
& & = \sum _n nt_n tr((-{\partial \over \partial \Lambda _{tr}} )^{k-1}
\Lambda ^{p+n+k-1}) e^{-trV[\Lambda ]} + \nn \\
& & + \sum _{a+b=p+k-1} {\partial \over
\partial t_a} \left( tr((-{\partial \over \partial \Lambda _{tr}} )^{k-1}
\Lambda ^b) e^{-trV[\Lambda ]} \right) = \nn \\
& & = \left( \sum _n nt_n \tilde {\cal W}^{(k)}_{n+p} +
\sum _{a+b=p+k-1} {\partial \over \partial t_a} \tilde {\cal W}^{(k)}_{b+1-k}
\right) e^{-trV[\Lambda ]} \hbox{  .  }
\eea
Eqs.(13) and, especially, (14) are trivially derived. Eqs.(12) and (13) imply
that all the $\tilde {\cal W}^{(k)}$-operators are in fact proportional to
linear combinations of Virasoro operators ${\cal L} = \tilde {\cal W}^{(2)}$.
 This may
explain how in the continuum limit a single Ward identity (10) can give rise
to entire set of constraints with lower spins. However, this topic is beyond
the scope of this Letter.
The continuum limit of these eqs. may be studied along the lines of [7].

\section{Formulation in terms of [9,10]}

While from the point of view of derivations
it is illuminating first to study the simplest Ward identity (3) in the
``1-matrix component" of 2-matrix model, and then apply it to the (more
sophisticated) analysis of 2-matrix case, it is of course possible to treat the
resulting  ${\cal W}$-constraints (10) as Ward identities in the entire
2-matrix model, related to the following infinitesimal change of integration
variables

\bea
\delta X & = & \Lambda ^p\hbox{, }    p \geq  0 \\
\delta \Lambda & = & \left( \sum_{m=0}^K ms_m \sum_{k=0}^{m-2} (-)^{k+1}
(V')^k X^{m-2-k} \right) \Lambda ^p + \hbox{"quantum    }
\hbox{  corrections"} \nn
\hbox{  .  }
\eea
This variation of variables induces the variation of potential:

\beq
\delta S = \left( \sum_{m=0}^K ms_m (-)^m
(V')^{m-1} \right) \Lambda ^p + \Lambda^{p+1} + \hbox{"quantum   }
\hbox{  corrections"}
\hbox{  .  }
\eeq
The first term in this expression gives rise to the "classical" part
of $\tilde {\cal W}$-algebra and the second one produces the derivative
$\partial / {\partial t_{p+1}}$ in (10).

While the $X$-component of the variation (16) (which is of course nothing
but (3)) does not change the integration measure  $DXD\Lambda $, this is not
true for  $\Lambda $-component. The corresponding Jacobian is responsible for
the ``quantum" contributions to (16) and (17).

Eqs. (16)-(17) provide a possible generalization to the 2-matrix case of the
derivation [9,10] of the Virasoro constraints in the discrete 1-matrix model
from
the Ward identities associated with the shift  $M \rightarrow  M +
\epsilon M^{n+1}$ ($n \geq  -1$) of the integration variable.

\section{Some results about  $\tilde {\cal W}$-operators}

\subsection{A short summary}

  The purpose of this section is to begin the
systematic study of  $\tilde {\cal W}$-constraints.

First of all we should stress that eq.(11) defines only {\it positive $(p
\geq  1-K$}, to be exact) harmonics  $\tilde {\cal W}^{(K)}_p$ of
$\tilde {\cal W}^{(K)}$-operators. We do not address the structure of entire
$\tilde {\cal W}^{(K)}$-operators below.

Second, an important question is whether the set of
  $\tilde {\cal W}$-constraints
(10) is closed. We shall demonstrate in sect.4.3 below that this is indeed
the case. Namely it will be proved that

\beq
[\tilde {\cal W}^{(K)}_p,\tilde {\cal W}^{(K)}_q] \in
Span \tilde {\cal W}^{(K)}_r \hbox{  ,  }
\eeq
where the notation {\it Span} implies all possible linear combinations
with coefficients linearly depending of times and derivatives
and  $r \geq  1-K$  as long as  $p,q \geq  1-K$. (In order to complete the
proof one should also demonstrate that the commutators of
$\tilde {\cal W}^{(K)}_p$
with lower-spin operators  $\tilde {\cal W}^{(n)}$, $n<K$
 are appropriately adjusted.
This piece of the proof is omitted from sect.4 at the moment: we do not know
any elegant way to represent the entire algebra of  $\tilde
{\cal W}$-operators.) It
deserves noting that (18) remains a highly non-trivial property of
$\tilde {\cal W}$-algebra: by no means it follows from the fact that
$\tilde {\cal W}^{(K)}_p$ may be expressed through  ${\cal L}_q$ or
  ${\cal J}_q$. Moreover
$\tilde {\cal W}^{(K)}Z = 0$
does not imply that lower  $\tilde {\cal W}^{(n)}$, $n<K$
annihilate  $Z$, therefore (18) is absolutely crucial for the closeness of the
set of constraints (11). The fact that this set is indeed closed makes an
interesting exercise to observe the appearance of the entire tower of
$\tilde {\cal W}^{(n)}$-constraints
(with all spins  $n\leq K)$ in the continuum limit
from the single spin-$K$ constraint at the discrete level.

Third, just the same  $\tilde {\cal W}$-operators were found in [6]
in a somewhat
different context. In sect 4.5 we shall prove that they are really the same,
thus demonstrating a kind of universal nature of
$\tilde {\cal W}$-operators, at least,
in the framework of discrete matrix models.

Fourth, the fact that the commutator of  $\tilde {\cal W}^{(K)}$-operators
in (18) is
not just proportional to  $\tilde {\cal W}^{(2K-2)}_{p+q}$
demonstrates that  $
\oplus_{K}
\tilde {\cal W}^{(K)}$ {\it is not}
a Lie algebra (to make it similar to  ${\cal W}_\infty $,
at least, the basis should be changed). This makes  $\tilde {\cal W}$  even
more
similar to conventional  $\cal W$-algebras which are also non-linear and closed
as
soon as only operators  $\tilde {\cal W}^{(n)}$ of spins  $n\leq K$  are
considered.

Thus the subject of  $\tilde {\cal W}$-algebras
may deserve further investigation.

\subsection{Other formulations of $\tilde {\cal W}$-operators}

  Before we turn to the derivation of (18) let us present an explicit
expressions for the first members of the  $\tilde {\cal W}$-family

\beq
\tilde {\cal W}^{(2)} = \tilde {\cal L}_n =
\sum _{k>0}kt_k {\partial \over \partial t_{k+n}} +
\sum _{a+b=n}{\partial ^2\over \partial t_a\partial t_b}
\eeq

\bea
\tilde {\cal W}^{(3)}_{p-2} = \sum _{k,l>0}kt_klt_l
{\partial \over \partial t_{p+k+l-2}} +
\sum _{k>0}kt_k\sum _{a+b=k+p-2}{\partial ^2\over \partial t_{a}\partial t_b} +
\nn \\
+
\sum _{k>0}kt_k\sum _{a+b=p-1}{\partial ^2\over \partial t_a\partial t
_{k+b-1}} +
 \sum _{a+b+c=p-2}{\partial ^3\over \partial t_a\partial t_b\partial t_c} +
{p(p-1) \over 2} {\partial \over \partial t_p}
\eea
For comparison we write down conventional ${\cal W}$-operators:

\beq
{\cal W}^{(2)}_n = {\cal L}_n =
 \sum _{k>0}kt_k {\partial \over \partial t_{k+n}} + {1 \over 2}
\sum _{a+b=n}{\partial ^2\over \partial t_a\partial t_b}
\eeq

\bea
{\cal W}^{(3)}_p =
 3 \sum _{k,l>0}kt_klt_l {\partial \over \partial t_{p+k+l}} +
3 \sum _{k>0}kt_k\sum _{a+b=k+p}{\partial ^2\over \partial t_a%
\partial t_b} +
\sum _{a+b+c=p}{\partial ^3\over \partial t_a\partial t_b\partial t_c}
\eea

These operators can be expressed in terms of free field.
Let us introduce also the corresponding current (its negative modes are
described by eq.(14)):

\beq
{\cal J}(z) = \sqrt{2}\partial \phi (z) = \sum _{k>0}kt_kz^{k-1} +
\sum _{k\geq 0}{1\over z^{k+1}} {\partial \over \partial t_k} =
\sum _{k>0}{\cal J}_{-k}z^{k-1} + \sum _{k\geq 0}{\cal J}_kz^{-k-1}
\eeq
Now ${\cal W}$-operators can be described as

\beq
{\cal W}^K_p = :{\cal J}^K:_{-K-3}
\eeq
where standard normal ordering :$\ldots$ : implies that all positive current
modes
should be pulled to the left and commutators should be thrown out.

Alternatively, standard normal ordering one can introduce another one (we shall
denote it as $\ddag$ $\ldots$ $\ddag$ ) which permits one to reproduce
$\tilde {\cal W}$-operators from the same formulas (24). Namely, new
ordering implies that the terms with ``wrong" order of current modes should be
just thrown out. Let us consider two simplest examples. The first is the
Virasoro
case, where positive generators can be written as a sum of three terms:

\beq
{\cal J}_+{\cal J}_- + {\cal J}_-{\cal J}_+ + {\cal J}_-{\cal J}_-.
\eeq
Then the standard normal ordering leads to the expression (21):

\beq
{\cal L} = 2{\cal J}_+{\cal J}_- + {\cal J}_-{\cal J}_-,
\eeq
but new ordering gives (23):

\beq
\tilde {\cal L} = {\cal J}_+{\cal J}_- + {\cal J}_-{\cal J}_-.
\eeq
Here it
is certainly possible to turn these expressions into each other by trivial
rescaling of time variables. But this is already not the case when considering
$\tilde {\cal W}^{(3)}$-generators.

Indeed, the proper sum of currents in the $\tilde {\cal W}^3$-case is:

\beq
({\cal J}_+{\cal J}_-{\cal J}_- + {\cal J}_-{\cal J}_+{\cal J}_- +
{\cal J}_-{\cal J}_-{\cal J}_+) + ({\cal J}_-{\cal J}_+{\cal J}_+ +
{\cal J}_+{\cal J}_-{\cal J}_+ + {\cal J}_+{\cal J}_+{\cal J}_-) +
{\cal J}_-{\cal J}_-{\cal J}_-.
\eeq
Then terms in brackets are equal to each other under standard normal ordering
and we obtain the formula (22).

In the case of $\tilde {\cal W}^{(3)}$-generators in the second brackets
the only term remains
and the first two terms should be taken from the first brackets.
But the term ${\cal J}_-{\cal J}_+{\cal J}_-$ should be provided by
additional restriction of the modes of bilinear combination of the right two
currents
$({\cal J}_+{\cal J}_-)_p$ to
$p\geq -1$. Then this term corresponds to right ordering and leads to
the second term in (20) and we obtain all
items with unit coefficients as in (20).
Thus we can rewrite the new ordering as follows:

\beq
\tilde {\cal W}^{(K)}_p = \ddag (\partial \phi )^K\ddag _p =
\ddag {\cal J}^K\ddag _p =
\sum_{\begin{array}{c}
p_1 + \ldots + p_K = p \\ p_K \ge 0 \\ p_{K-1} + p_K \ge -1 \\ p_{K-2}
 + p_{K-1} + p_K \ge -2 \\ \ldots \end{array}}
  {\cal J}_{p_1}...{\cal J}_{p_K}\hbox{,
with }\ p\geq 1-K.
\eeq
Its characteristic property is that  $\ddag AB\ddag  =
\ddag A(\ddag B\ddag )\ddag $ , but $\ddag AB\ddag  \neq
\ddag (\ddag A\ddag )B\ddag $.

The last comment concerns the possibility to reformulate the actual constraints
(10) in terms of pure $\tilde {\cal W}^{(K)}$-operators by appropriate shift
of time variables. While this is trivial to do in the case of $K=2:$

\beq
\tilde {\cal W}^{(2)}_n \{ t_k \} - {1\over 2s_2} {\partial \over \partial
t_{n+2}} =
\tilde {\cal W}^{(2)}_n \{ t_k  - {1 \over 4s_k} \delta _{k,2} \}
\eeq
things become worse already for $K=3$. Namely, even to absorb derivative
$\partial /\partial t_{p+3}$ into  $\tilde {\cal W}^{(3)}_p$ one needs to
perform a complicated recursive redefinition of times:

\beq
\tilde {t}_k = t_k + \Delta t_k \hbox{  ,  }
\eeq

\bea
& & 3s_3\tilde {\cal W}^{(3)}_{p-2}\tilde{t} + 2s_2{\cal
W}^{(2)}_{p-1}\tilde{t}
=
\nn \\
& = & 3s_3\tilde {\cal W}^{(3)}_{p-2}\{t\} + 2s_2{\cal W}^{(2)}_{p-1}\{t\} +
3s_3\sum  _k k\Delta t_k\tilde {\cal W}^{(2)}_{p+k-2}\{t\} +
\nn \\
& + & \left\lbrace 3s_3\sum _{k+l=n+2}k(\Delta t_k)l(\Delta t_l) +
2s_2(n+1+\Delta t_{n+1})\right\rbrace {\partial \over \partial t_{p+n}}
\hbox{  .  }
\eea

Thus it is clear that the most we can achieve is to adjust $\{\Delta t_k\}$ in
such a way, that the expression in braces in the last term is equal $to
-\delta _{n,1}$. Such adjusting is always possible, at least, in the iteration
form

\bea
\Delta t_1 & = & 0 \hbox{  ,  }
\nn \\
\Delta t_2 & = & - 1/4s_2 \hbox{  ,  }
\nn \\
\Delta t_{n+1} & = & - {3s_3\over 2(n+1)s_2}
\sum ^n_{k=2}k(n+2-k)\Delta t_k\Delta t_{n+2-k}
\nn \\
\hbox{for   }  n & > & 1 \hbox{  ,  }
\eea
but it does not seem too illuminating.

\subsection{Computation of $[\tilde {\cal W}^{(K)}_p,\tilde {\cal W}^{(K)}_q]$}

To evaluate this commutator we shall make direct use of the definition (11) of
$\tilde {\cal W}$-operators and note that $\Lambda $ and $t_k$ are independent
(and, therefore, commuting) variables. Thus

\bea
\tilde {\cal W}^{(K+1)}_p \tilde {\cal W}^{(K+1)}_q e^{-trV} & = &
\tilde {\cal W}^{(K+1)}_p tr\left\lbrace (-
{\partial \over \partial \Lambda _{tr}})^K \Lambda ^{q+K}\right\rbrace
e^{-trV} = \nn \\
& = & tr\left\lbrace (- {\partial \over \partial \Lambda _{tr}})^K
\Lambda ^{q+K}\right\rbrace  \tilde {\cal W}^{(K+1)}_p e^{-trV} = \nn \\
& = & tr\left\lbrace (- {\partial \over \partial \Lambda _{tr}})^K
\Lambda ^{q+K}\right\rbrace  tr\left\lbrace (-
{\partial \over \partial \Lambda _{tr}})^k \Lambda ^{p+K}\right\rbrace
e^{-trV},
\eea
so that

\beq
[\tilde {\cal W}^{(K+1)}_p\hbox{, } \tilde {\cal W}^{(K+1)}_q] e^{-trV} =
\left[ tr\left\lbrace (- {\partial \over \partial \Lambda _{tr}})^K
\Lambda ^{q+K}\right\rbrace \hbox{, } tr\left\lbrace (-
{\partial \over \partial \Lambda _{tr}})^K \Lambda ^{p+K}\right\rbrace \right]
e^{-trV}
\eeq
(to simplify the formulae below we consider $\tilde {\cal W}^{(K+1)}_p$ instead
of $\tilde {\cal W}^{(K)}_p).$

Our strategy in the derivation of (18) from the $r.h.s$. of (35) is as follows:
\bigskip

({\it i}) Let us carry the second $(-
{\partial / {\partial \Lambda _{tr}}})^K$ at the $r.h.s$. of (35) to the
{\it left}. Neglecting all the commutators of $(-
{\partial / {\partial \Lambda _{tr}}})^K$ with $\Lambda ^{q+K}$, one obtains

\beq
(- {\partial \over \partial \Lambda _{tr}})^K_{\alpha \beta }(-
{\partial \over \partial \Lambda _{tr}})^K_{\gamma \delta }(\Lambda ^{q+K})_{%
\beta \alpha }(\Lambda ^{p+K})_{\delta \gamma } e^{-trV},
\eeq
which cancels in the commutator (35). Therefore,
only the terms
$\left[ (\Lambda ^{q+K})_{\beta \alpha } \hbox{  ,  }
(-
{\partial / {\partial \Lambda _{tr}}})^K_{\gamma \delta }\right] $
from the commutator are of
interest, and we can consider

\beq
(-
{\partial \over \partial \Lambda _{tr}})^K_{\alpha \beta }
\left\lbrace \left[ (%
\Lambda ^{q+K})_{\beta \alpha }\ ,\ (-
{\partial \over \partial
\Lambda _{tr}})^K_{\gamma \delta }\right] (\Lambda ^{p%
+K})_{\delta \gamma }\right\rbrace  e^{-trV} - (p \leftarrow \rightarrow q)
\eeq
instead of the $r.h.s$. of (35).
\bigskip

({\it ii}) Expression in braces in (37) contains at most $K-1$ derivatives. Let
us pull all these derivatives to the {\it right} and then use the relation

\beq
(- {\partial \over \partial \Lambda _{tr}})_{\mu \nu } e^{-trV} =
(V')_{\mu \nu } e^{-trV} = \sum  _k kt_k (\Lambda ^{k-1})_{\mu \nu } e^{-trV}.
\eeq
Then the factor $kt_k$ can be pulled to the very {\it left},
since $t_k$'s commute
with all $\Lambda $'s and $\partial /\partial \Lambda _{tr}$'s.
\bigskip

({\it iii}) What remains now from expressions in braces in (35)
is a combination
of powers of $\Lambda $'s which in fact has only two free indices
$\beta \alpha $. Therefore, each item can be written as
$(\Lambda ^a)_{\beta \alpha }$ with some power {\it a} multiplied by some
contribution of $tr\Lambda ^b$ with various powers of $b$. All these traces can
be shiftted to the {\it right}. Then we use the fact that

\beq
(tr\Lambda ^b) e^{-trV} = - {\partial \over \partial t_b} e^{-trV}.
\eeq
Again $\partial /\partial t_b$ can be carried to the left of all $\Lambda 's$
and $\partial /\partial \Lambda _{tr}'s$ (but to the right of $t_k's$ which are
arising at the step ({\it ii})).
\bigskip

({\it iv}) Thus we got rid of all contributions like
$\partial /\partial \Lambda _{tr}$ and $tr\Lambda ^b$ to the braces in (37).
What remains is (a linear combination with ``coefficients" made from $kt_k$ and
$\partial /\partial t_b$ of)

\beq
(- {\partial \over \partial \Lambda _{tr}})^K_{\alpha \beta }
\Lambda ^a_{\beta \alpha } e^{-trV} = \tilde {\cal W}^{(K+1)}_{a-K} e^{-trV}.
\eeq
This is just the statement (18).

\subsection{Two manifest examples}

In order to illustrate this somewhat abstract consideration we present now two
more explicit examples. The first one concerns Virasoro
operators $\tilde {\cal W}^{(2)}$
--- it is a quite trivial case. The expression from braces in (37) is now (note
that $({\partial / {\partial \Lambda _{tr}}})_{\alpha \beta } =
({\partial / {\partial \Lambda }})_{\beta \alpha }$):

\beq
\left[ (\Lambda ^{q+1})_{\beta \alpha },(-
{\partial \over \partial \Lambda _{tr}})_{\gamma \delta }\right] (\Lambda ^{p+1
})_{\delta \gamma } =
\sum _{a+b=q}(\Lambda ^a)_{\beta \delta }(\Lambda ^b)_{\gamma \alpha }(\Lambda
^{p+1})_{\delta \gamma } = (q+1)(\Lambda ^{p+q+1})_{\beta \alpha },
\eeq
and the $r.h.s$. of (35) becomes

\beq
[(q+1) - (p+1)] tr{\partial \over \partial \Lambda _{tr}} \Lambda ^{p+q+1}
e^{-trV} = (p-q) \tilde {\cal W}^{(2)}_{p+q} e^{-trV},
\eeq

$i.e.$

\beq
[\tilde {\cal W}^{(2)}_p\hbox{, } \tilde {\cal W}^{(2)}_q] = (p-q)
\tilde {\cal W}^{(2)}_{p+q}\hbox{, }   p,q\geq -1,
\eeq
as it was expected. This example is not very representative as the steps
({\it ii}) and ({\it iii}) are in fact absent.

Another example of $\tilde {\cal W}^{(3)}$-operators involves all the steps
({\it i})-({\it iv}). Expression in braces in (37) is now (note that
$({\partial / {\partial \Lambda _{tr}}})^2_{\alpha \beta } =
({\partial / {\partial \Lambda }})_{\epsilon \alpha }
({\partial / {\partial \Lambda }})_{\beta \epsilon }$):

\bea
& - & \left[ (\Lambda ^{q+2})_{\beta \alpha },({\partial
\over \partial \Lambda _{tr}
})^2_{\gamma \delta }\right] (\Lambda ^{p+2})_{\delta \gamma } = \nn \\
& = & - \sum _{a+b+c=q}\left[ (\Lambda ^a)_{\beta \epsilon }(\Lambda ^b)_{%
\gamma \delta }(\Lambda ^c)_{\epsilon \alpha } +
(\Lambda ^a)_{\beta \delta }(\Lambda ^b)_{\epsilon \epsilon }(\Lambda ^c)_{%
\gamma \alpha }\right] (\Lambda ^{p+2})_{\delta \gamma } + \nn \\
& + & \sum _{a+b=q+1}\left[ (\Lambda ^a)_{\beta \epsilon }(\Lambda ^b)_{%
\gamma \alpha }({\partial \over \partial \Lambda })_{\delta \epsilon } +
(\Lambda ^a)_{\beta \delta }(\Lambda ^b)_{\epsilon \alpha }({\partial \over %
\partial \Lambda })_{\epsilon \gamma }\right] (\Lambda ^{p+2})_{\delta \gamma
}%
{}.
\eea
In order to complete the step ({\it ii}) we should carry the remaining
derivatives $\partial /\partial \Lambda $ in the second sum to the right, so
that the $r.h.s$. of (44) transforms into

\bea
& - & \sum _{a+b+c=q}\left[ (\Lambda ^{a+c})_{\beta \alpha }
tr\Lambda ^{b+p+2} +
(\Lambda ^{a+c+p+2})_{\beta \alpha } tr\Lambda ^b\right]  -
\nn \\
& - & \sum _{a+b=q+1} \sum _{c+d=p+1}\left[ (\Lambda ^{a+b+d})_{\beta \alpha }
tr\Lambda ^c + (\Lambda ^{a+b+c})_{\beta \alpha } tr\Lambda ^d\right]  +
\nn \\
& + & \sum _{a+b=q+1}\left[ (\Lambda ^a)_{\beta \epsilon }(\Lambda ^{b+p+2})_{%
\delta \alpha }({\partial \over \partial \Lambda _{tr}})_{\epsilon \delta } +
(\Lambda ^{a+p+2})_{\beta \gamma }(\Lambda ^b)_{\epsilon \alpha }({\partial %
\over \partial \Lambda _{tr}})_{\gamma \epsilon }\right] .
\eea
Using the (38) we can turn it into

\bea
& - & \sum _{a+b=q}(a+1)\left[ (\Lambda ^a)_{\beta \alpha } tr\Lambda ^{b+p+2}
+
(\Lambda ^{a+p+2})_{\beta \alpha } tr\Lambda ^b\right]  -
\nn \\
& - & 2(q+2) \sum _{a+b=p+1}(\Lambda ^{a+q+1})_{\beta \alpha } tr\Lambda ^b -
2(q+2) \sum  _k kt_k (\Lambda ^{p+q+k+2})_{\beta \alpha }.
\eea
Steps ({\it iii}) and ({\it iv}) are now trivial and we read off from (46):

\bea
[\tilde {\cal W}^{(3)}_p\hbox{, } \tilde {\cal W}^{(3)}_q] & = & 2(p-q) \sum
_k
kt_k \tilde {\cal W}^{(3)}_{p+q+k} - \nn \\
& - & \bigl \lbrace \sum _{a+b=q}(a+1)\left[
{\partial \over \partial t_{b+p+2}}
\tilde {\cal W}^{(3)}_{a-2} + {\partial \over \partial t_b}
\tilde {\cal W}^{(3)}_{a+p}\right]  + \nn \\
& + & 2(q+2)\sum _{a+b=p+1}{\partial \over \partial t_b}
\tilde {\cal W}^{(3)}_{a+q-1} - (p\leftarrow \rightarrow q)
\bigr \rbrace = \nn \\
& = & 2(p-q) \sum  _k kt_k \tilde {\cal W}^{(3)}_{p+q+k} + \nn \\
& + & \left[ \sum ^{p+1}_{a=0} (2p-q-2a) - \sum ^{q+1}_{a=0} (2q-p-2a)\right]
{\partial \over \partial t_b} \tilde {\cal W}^{(3)}_{p+q-a},   p,q\geq -2.
\eea
For $p>q$ the item in square brackets on the $r.h.s$. can be equivalently
rewritten as

\beq
3(p-1) \sum _{a+b=q+1}{\partial \over \partial t_a}
\tilde {\cal W}^{(3)}_{b+p-1} + \sum ^{p+1}_{a=q+2}(2p-q-2a)
{\partial \over \partial t_a} \tilde {\cal W}^{(3)}_{p+q-a}.
\eeq
\bigskip

For illustrative purposes let us show that operators
$
\tilde {\hbox{\kern-2pt\hbox{$\tilde {\cal W}$}}}^{(3)}_p \equiv
 \tilde {\cal W}^{(3)}_p + \alpha (\partial /\partial t_{p+3})$
also form a closed set of constraints for $p\geq -2$. In accordance with (47),
(48), it is enough to check that as soon as $p>q\geq -2$

\bea
\biggl[ {\partial \over \partial t_{p+3}}\hbox{ , }
\tilde {\cal W}^{(3)}_q\biggr]  - (p\leftarrow \rightarrow q) = 2(p-q) \sum  _k
kt_k {\partial \over \partial t_{p+q+3}} + \nn \\
+ 3(p-q) \sum _{a+b=q+1}{\partial \over \partial t_b}
{\partial \over \partial t_{a+p+3}} + \sum ^{p-q-1}_{b=0}(2p-3q-4-2b)
{\partial \over \partial t_{q+2+b}} {\partial \over \partial t_{p+1-b}} .
\eea
The $l.h.s$. of this relation is easily evaluated with the help of explicit
formula (20) for $\tilde {\cal W}^{(3)}_q$ and is equal to

\bea
2(p-q) \sum  _k kt_k {\partial \over \partial t_{p+q+3}} + (p-q)
\sum _{a+b=p+q+3}{\partial ^2\over \partial t_a\partial t_b} + \nn \\
+ (p+3) \sum _{a+b=q+1}{\partial ^2\over \partial t_{a+p+2}\partial t_b} -
(q+3) \sum _{a+b=p+1}{\partial ^2\over \partial t_{a+q+2}\partial t_b}\hbox{ .}
\eea
It is a trivial arithmetic exercise to check that (50) coincides with the
$r.h.s$. of (49) as soon as $p>q\geq -2.$

\subsection{$\tilde {\cal W}$-constraints in Kontsevich's like models}

  In order to add
new colours to the picture of  $\tilde {\cal W}$-constraints let us
present some
details concerning their appearance in the framework of 1-matrix Kontsevich's
models.
Indeed, as shown in [6] the Ward identity (2) which has been applied
above to the derivation of  $\tilde {\cal W}^{(K)}$-constraints in 2-matrix
model may
be represented in a form of  $\tilde {\cal W}^{(K-1)}$-constraint imposed on
the
partition function of 1-matrix model (1) itself. Following sect. 3.2 of
ref.[6] we shall consider here only a simplified example, or what was
called there a discrete counterpart of  $\tilde {\cal W}$-constraint in the
model
(1). Namely (see eq.(26) of ref.[6])

\beq
tr[M^q({\partial \over \partial M_{tr}})^K]{\cal F}\{t_k\} =
\sum _{p\geq 1-K}tr\ M^{q-p-K} \tilde W^{(K)}{\cal F}\{t_k\} \hbox{  .  }
\eeq
for any function  ${\cal F}\{t_k\}$  with the arguments

\beq
t_k = {1\over k} trM^{-k} \hbox{  ,  }
\eeq

\beq
t_0 = - tr\hbox{ logM} \hbox{  .  }
\eeq
We shall demonstrate now that  $\tilde {\cal W}^{(K)}$-operators in the
$r.h.s$.
of
(51) are exactly the same as  $\tilde {\cal W}^{(K)}$'s
of this paper defined through
(11)

\beq
tr[(- {\partial \over \partial \Lambda _{tr}})^{K-1}\Lambda ^{p+K-1}]
e^{-trV\{%
\Lambda )} = \tilde W^{(K)}_pe^{-trV\{\Lambda \}} \hbox{  .  }
\eeq
In order to prove the equivalence of  $\tilde {\cal W}$-operators in (51) and
(54) we
take  ${\cal F}\{t_k\} = e^{trV}$, where

\beq
trV = \sum    t_k tr\Lambda ^k = - tr\hbox{ logM }\ trI +
\sum _{k\geq 1}{1\over k} trM^{-k} tr\Lambda ^k
\hbox{  ,  }
\eeq
so that

\beq
e^{-trV} = \det (M \otimes  I - I \otimes  \Lambda )^{-1} \hbox{  .  }
\eeq
Then it remains to prove that (56) satisfies

\beq
tr[M^q({\partial \over \partial M_{tr}})^K]e^{-trV} =
 \sum _{p\geq 1-K}tr\ M^{q-p-K}
tr[(-{\partial \over \partial \Lambda _{tr}})^{K-1}\Lambda ^{p+K-1}]e^{-trV\{%
\Lambda )}
\eeq
The $r.h.s$. of (57) may be further simplified: the sum over  $p$  is easily
evaluated and gives

\beq
tr(M^q \otimes  I)(I \otimes (- {\partial \over \partial \Lambda _{tr}})^{K-1})
{e^{-trV}\over M \otimes I - I \otimes  \Lambda } =
tr[M^q({\partial \over \partial M_{tr}})]tr({\partial
\over \partial M_{tr}})^{%
K-1} e^{-trV}
\eeq
so that (57) becomes a trivial identity

\beq
tr\left( M^q {\partial \over \partial M_{tr}} \otimes
I\right) \left( ({\partial \over \partial M_{tr}})^{K-1} \otimes  I + I
\otimes  (-{\partial \over \partial \Lambda _{tr}})^{K-1}\right)
{1\over \det (M \otimes I  - I \otimes  \Lambda )} \equiv  0
\eeq
To make this proof a bit more transparent let us repeat it for the case of  $N
= 1$  (just numbers  $m$  and  $\lambda $  instead of matrices  $M$  and
$\Lambda $):

\beq
m^q({\partial \over \partial m})^K {1\over m-\lambda } =
\sum _{p\geq 1-K}m^{q-p-K}(-{\partial \over \partial \lambda })^{K-1}
\lambda ^{p+K-1} {1\over m-\lambda } =
 m^q (-{\partial \over \partial \lambda })^{K-1} {1\over (m-\lambda )^2}
\eeq
Both sides of this relation are equal to  $(K-1)!
{\displaystyle{m^q \over (m-\lambda )^K}}$
and, thus, coincide.

A really interesting question about Kontsevich's-like models (1) concerns
the reformulation of actual Ward identities (2) in terms of appropriate time
variables  $T^{(K)}_k$, (in the case of  $K = 2\ \ T^{(2)}_k$ are defined by
(3) instead of (52)-(53)). This is however beyond the scope of the present
paper: the purpose of this section was to demonstrate that
$\tilde {\cal W}$-operators arise at least in two different contexts.

\section{Conclusion}

To conclude, we demonstrated that the Ward-identities of Kontsevich-like
models, derived in [6], are enough to obtain a closed (and presumably)
complete set of Ward identities (loop equations) in discrete 2-Hermitean
matrix model. These loop-equations involve $\tilde {\cal W}$-operators
acting on one of potentials in the 2-matrix model, while the "highest"
spin involved in these operators coincides with the power of another potential.
\footnote{So one can look at these constraints as at Ward identities of proper
1-matrix model (in terms of variable $\Lambda $). It differs from
"true" 2-matrix model $\cal W$-constraints depending on two sets of times,
see the example of this in [11].}
This statement is the discrete
counterpart of the suggestion of Gava and Narain [7], concerning the
"asymmetric" (i.e. two potentials do not coincide) continuum limit of
2-matrix model, which is probably different from the symmetric limit,
originally examined in [2].

We demonstrated also that these unconventional $\tilde {\cal W}$-operators
arise at least at two slightly different places. First, if we
study an analogue of Kontsevich's model with potential $W[X]$ of $M$-th
power (or rather what was called in [6] its discrete "analogue"),
then operators $\tilde {\cal W}^{(M-1)}$ arise. Second, if
 just the same $X$-integral is
treated as the constituent of 2-matrix model, then operators
$\tilde {\cal W}^{(M)}$
arise.
Note that in these two examples the
"spins" of $\tilde {\cal W}$-operators
are different. Of course, their meanings are
different too, as well as the time-variables, through which they are
expressed: these are really two {\it different} examples where {\it the same}
$\tilde {\cal W}$-operators seem to arise.

\bigskip

 Detailed study of $\tilde {\cal W}$-algebras, their relations to
 conventional $\cal W$-algebras,
their continuum limits (in the matrix-model sense) etc. deserve further
analysis, which is beyond the scope of this Letter.

\bigskip
\bigskip
\begin{center}{{\Large\bf Acknowledgements}}\end{center}
\medskip

We benefited a lot from illuminating discussions on the subject with
E.Corrigan, E.Gava, K.Narain, A.Niemi.

We are grateful to the hospitality of the Research Institute for Theoretical
Physics (TFT)
at Helsinki University where some pieces of this work have been done. A.Mor. is
also very much indebted for the hospitality of Dept. of Math. Sci. of Durham
University. The work of A.Mar. and A.Mir. was partially supported by the
Re\-se\-a\-rch In\-s\-ti\-tu\-te for The\-o\-re\-ti\-cal Physics (TFT) at
Hel\-sin\-ki Uni\-ver\-si\-ty while that of A.Mor. by the UK Science
 and En\-ge\-ne\-er\-ing
Co\-un\-cil through the Vi\-si\-ting Fel\-low\-ship prog\-ram and by
NO\-R\-DI\-TA.

\bigskip
\bigskip
\begin{center}{{\Large\bf References}}\end{center}
\medskip
1. V.Kazakov  Mod.Phys.Lett. A4 (1989) 2125

  E.Brezin, V.Kazakov   Phys.Lett. B236 (1990) 144

    M.Douglas, S.Shenker  Nucl.Phys. B335 (1990) 635

    D.Gross, A.Migdal  Phys.Rev.Lett. 64 (1990) 127\\
2. M.Fukuma, H.Kawai, R.Nakayama  Int.J.Mod.Phys. A6 (1991) 1385

  R.Dijkgraaf, E.Verlinde, H.Verlinde  Nucl.Phys. B356 (1991) 574\\
3. A.Gerasimov et.al. Nucl.Phys. B357 (1991) 565\\
4. Yu.Makeenko et.al. Nucl.Phys. B356(1991) 574\\
5. M.Kontsevich  Funk.Anal. i Priloz. 25 (1991) 50\\
6. A.Marshakov et al. "On equivalence of topological and
quantum $2d$ gravity", Phys.Lett. B, in press\\
7. E.Gava, K.Narain  Phys.Lett. B263 (1991) 213\\
8. M.L.Mehta Comm.Math.Phys. 79 (1981) 327\\
9. A.Mironov, A.Morozov Phys.Lett. B252 (1990) 47\\
10. J.Ambjorn, J.Jurkiewicz, Yu.Makeenko  Phys.Lett. B251 (1990) 517\\
11. A.Marshakov, A.Mironov, A.Morozov Phys.Lett. B265 (1991) 99\\

\end{document}